\documentclass[prb,twocolumn,amsfonts,superscriptaddress,showpacs]{revtex4}
 
\usepackage[]{graphicx} 
\usepackage{amsbsy} 
\usepackage{float}
\usepackage{color}
\usepackage{subfigure}

\usepackage{amsmath}
\usepackage{braket} 
 
\begin{document} 
\title{Electronic structure  of  CeRu$_4$Sn$_6$:\\ a density functional  plus dynamical mean field theory study}
\author{P.~Wissgott and K.~Held}
\affiliation{Institute for Solid State Physics, Vienna University of Technology, AT-1040 Vienna, Austria}
\date{\today} 
\begin{abstract} 
The Kondo system CeRu$_4$Sn$_6$ shows a strong anisotropy in its electric, optic and magnetic properties. We employ density functional theory plus dynamical mean field theory  and show  that the predominant Ce-$f$ state has total angular moment $J=5/2$ and $z$-component $m_J=\pm 1/2$ in agreement with recent X-ray absorption experiments.  Even though   CeRu$_4$Sn$_6$ has the direct gap of a Kondo insulator through most of the Brillouin zone it remains weakly metallic. This is because of  (i) a band crossing in the $z$-direction and  (ii) a  negative indirect gap.
\end{abstract}
\date{\today}

\maketitle

In $f$-electron systems such as CeRu$_4$Sn$_6$ we have the generic situation
that  narrow, strongly interacting $f$ bands hybridize with weakly correlated
conduction bands,  as  exemplified by the periodic Anderson model.
This gives rise to a hybridization gap at the crossing of 
$f$ and conduction band, already without $f$-$f$ interaction  e.g.\ in  density functional theory (DFT). Electronic correlations strongly renormalize this gap, resulting in a gapped Kondo resonance which can be understood in a quasiparticle picture. 
If the gap is at the Fermi level
 one has a Kondo insulator; otherwise one has a metal with a typically heavy effective mass \cite{Ste84.1,Col07.1}.
 Even more complicated is the situation in anisotropic Kondo insulators such as CeNiSn \cite{Tak90.1}. Here, the  physical properties such as the susceptibility and the conductivity are strongly anisotropic. One possible explanation is 
 a hybridization gap with nodes \cite{Ike96.1,Mor00.1,Yam12.1} so that 
in some directions the bands cross the Fermi level whereas there is a gap in other parts of the Brillouin zone. Besides, also a V-shaped DOS \cite{Kyo90.1} or extrinsic effects such as off stoichiometry, impurities\cite{Sch92.1} or topological surfaces states have been considered as a microscopic origin of the residual metallicity in some directions.

\begin{figure}[t]
\includegraphics[height=2.cm,clip=,angle=0]{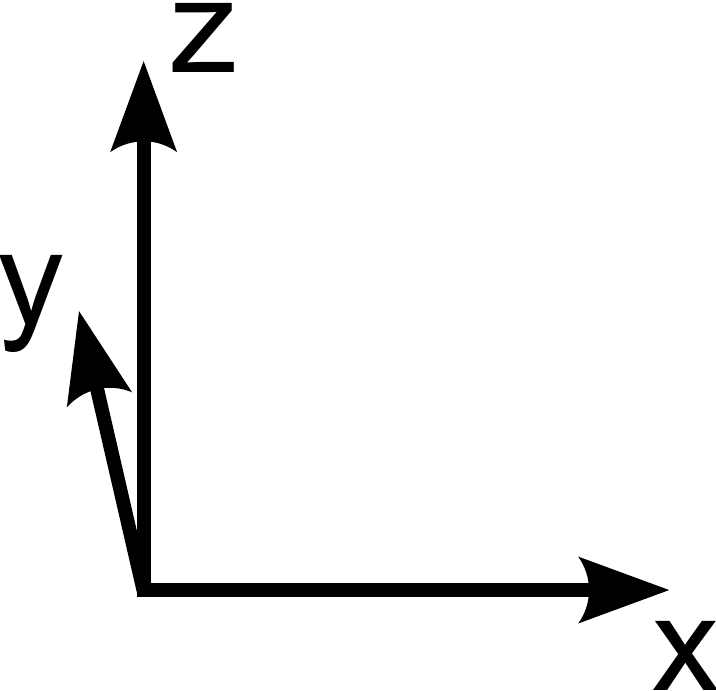}
\includegraphics[height=6.cm,clip=,angle=0]{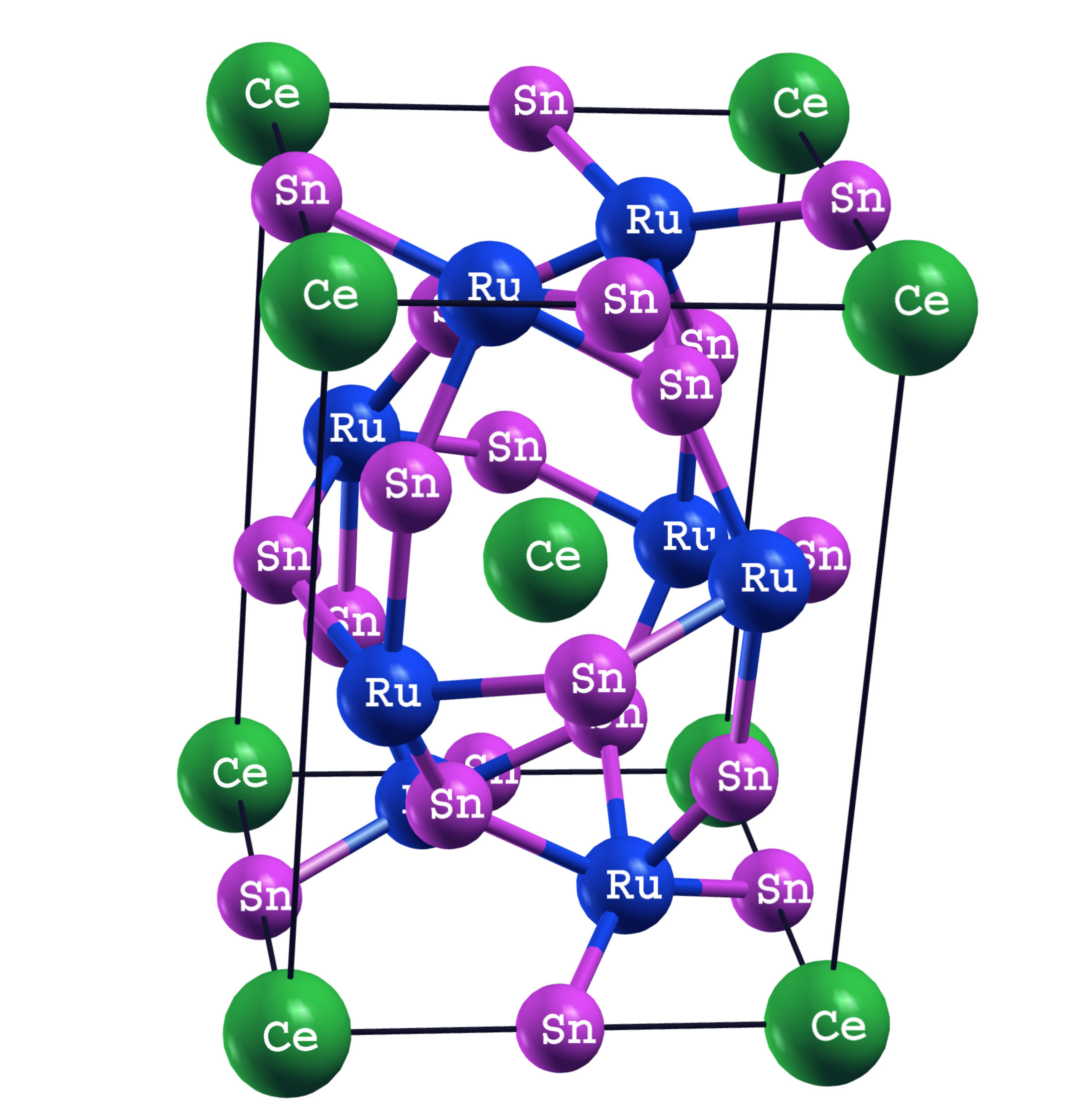}
    \caption{Crystal structure of CeRu$_4$Sn$_6$.}
    \label{Fig:CeRu4Sn6_crystalstructure1}
\end{figure}

Such an anisotropy has also been observed
in 
CeRu$_4$Sn$_6$ single crystals  \cite{Pas10.1} with a simpler,  tetragonal
 I$\bar 4$2m crystal structure\cite{Ven90.1, Das92.1}  (Fig.~\ref{Fig:CeRu4Sn6_crystalstructure1}): 
the optical conductivity shows a weak Drude-like feature  in the  $a-a$ plane  whereas it has a dip at low frequencies
in the $c$-direction \cite{Guritanu2011}. This has been confirmed by
density functional theory plus dynamical mean field theory (DFT+DMFT) \cite{Kotliar2006,Held2007} calculations  of the optical conductivity \cite{Guritanu2011}. Likewise the thermopower \cite{Strydom2005,Hanel14}, resistivity \cite{Winkler2011} and magnetic properties \cite{Paschen2010} are strongly anisotropic.

In the present paper, we discuss the   DFT+DMFT calculations in more detail. In particular we show that electronic correlations and the Kondo effect reverse the crystal field splitting between $J=5/2$ $|m_J|=1/2$ and $J=5/2$ $|m_J|=3/2$ Ce-$f$ states. Here, $J$ is the  total angular moment and  $m_J$ its $z$-component. Consequently the $|m_J|=1/2$ states get predominantly occupied with decreasing temperature (increasing correlation effects). Nonetheless, the evolving Kondo resonance at the Fermi level also  contains    $|m_J|=3/2$ states. This DFT+DMFT finding agrees with recent x-ray experiments \cite{Severing} which  best fit to $|m_J|=1/2$.
We also discuss why CeRu$_4$Sn$_6$ remains metallic.
In section \ref{Sec:DFT} we present the DFT bandstructure and Wannier orbitals. In Section \ref{Sec:DMFT} the orbitally resolved DMFT spectra and self energies and their temperature-dependence is discussed. Finally, Section \ref{Sec:conclusion} summarizes our  results.

\section{Density functional theory}
\label{Sec:DFT}
Starting point is the experimental  tetragonal \emph{I$\bar 4$2m} structure of
CeRu$_4$Sn$_6$  with lattice parameters $a=6.8810$~\AA, and $c=9.7520$~\AA.\cite{Poettgen97} Here, the Ce atoms are surrounded by a "cage"  with four nearest neighbor  Ru atoms and four next nearest neighbor  Sn atoms which are only $2$\%  further away, see Fig.~\ref{Fig:CeRu4Sn6_crystalstructure1}. 
Our first step is to calculate the  DFT electronic structure with Wien2K\cite{Blaha1990} using  the PBE-GGA potential and  a fine $k$-mesh of $10000$ points to obtain accurate results. Spin-orbit effects are important due to the heavy atoms  and  hence included. 

\begin{figure}[t]
\includegraphics[width=8.cm,clip=,angle=0]{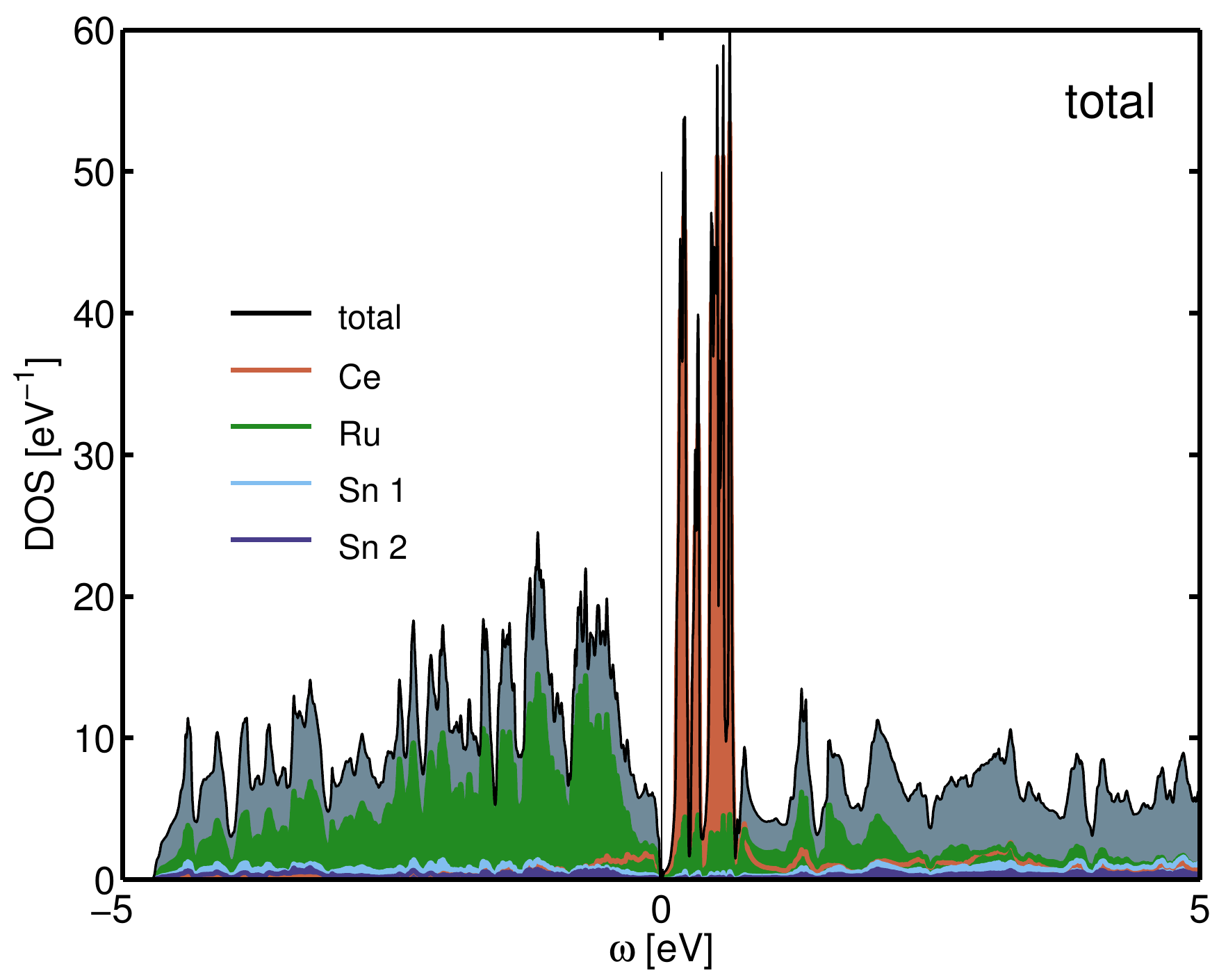}

\includegraphics[width=8cm,clip=,angle=0]{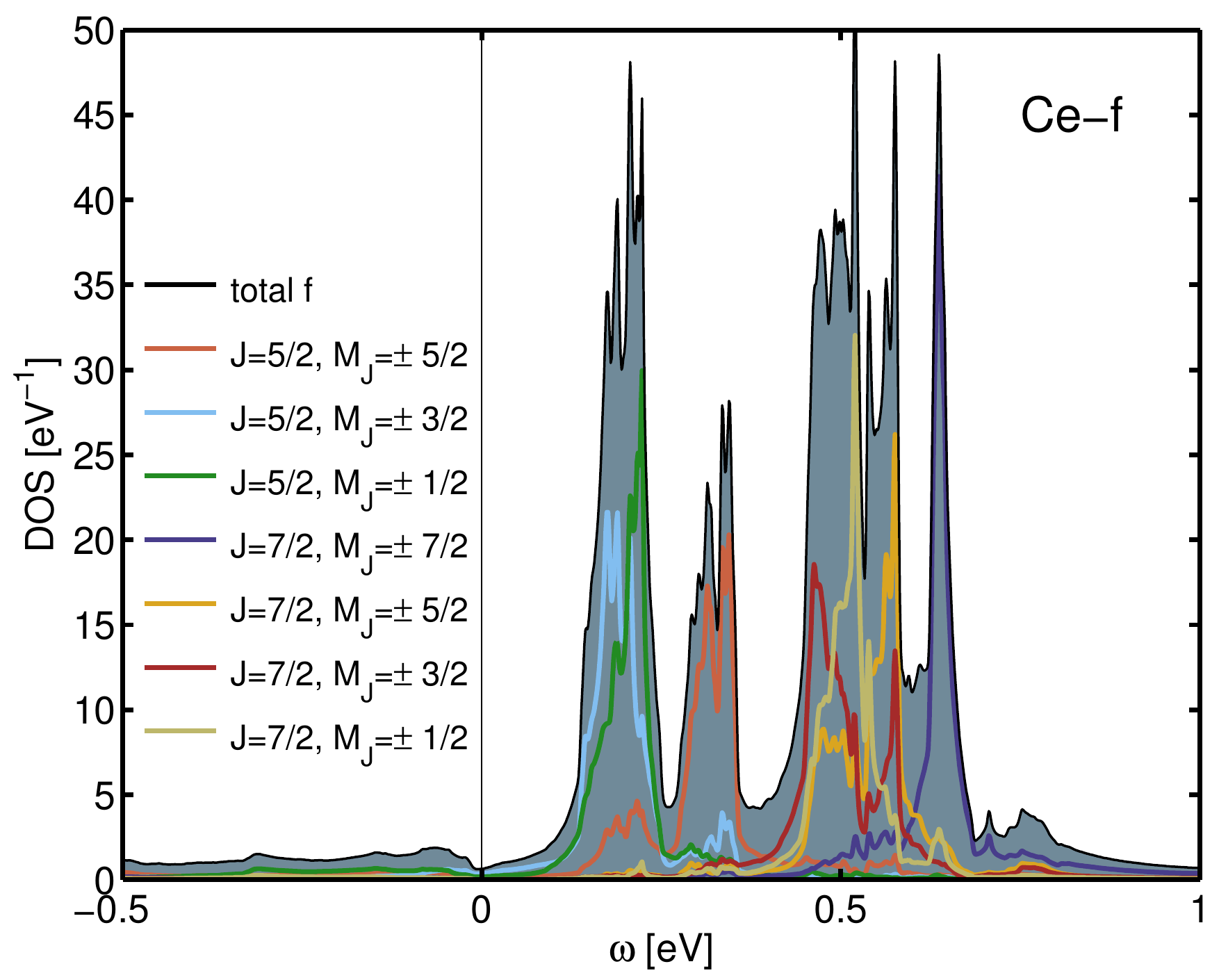}
    \caption{Top: DFT partial density of states resolving the most important atom-wise contributions: Ce-$f$, Ru-$d$ and Sn-$p$. Bottom: Zoom in and further differentiation of the f-states in the $(J,m_J)$-basis.}
    \label{Fig:CeRu4Sn6_DOS1}
\end{figure}

The DFT partial densities of states for CeRu$_4$Sn$_6$ are shown in Fig.~\ref{Fig:CeRu4Sn6_DOS1}. The calculation reveals a mixed manifold around the Fermi level with main contributions of Ce-$f$, Ru-$d$ and Sn-$p$ character \cite{footnote}. The narrow  $f$-bands are essentially  placed between 0.1 and 0.7 eV  above the Fermi level in DFT. Because of the small hybridization tail below the Fermi level,  the DFT  $f$-filling is nonetheless surprisingly large, i.e., $0.7$ electrons per Ce atom.  

Fig.~\ref{Fig:CeRu4Sn6_DOS1} (bottom) shows that the
spin-orbit coupling separates the Ce-$f$ states into  $J=5/2$ states, whose center of gravity is approximately around $0.2$ eV, and  $J=7/2$ states around $0.5$ eV. The crystal field further splits  the $J=5/2$ manifold into  $m_J=\pm 5/2$ character at higher energy and $m_J=\pm 3/2$ as well as $m_J=\pm 1/2$ states at essentially the same energy.  The crystal field splitting between the latter is  only $\sim 0.05\,$eV with   $m_J=\pm 3/2$ 
below  $m_J=\pm 1/2$. This reflects  in the 
 small splitting in
 the $f$-level at $\sim 0.2\,$eV in
Fig.~\ref{Fig:CeRu4Sn6_DOS1} (bottom).
Nonetheless, the largest contribution below the Fermi level has $(J=5/2,m_J=1/2)$ character because these states hybridize more strongly with Ru-$d$ than the other $f$ orbitals [the integrated weight below the Fermi level is larger for the   $m_J=\pm 1/2$ orbital in Fig.~\ref{Fig:CeRu4Sn6_DOS1} (bottom)].
Apart from the $f$ partial density of states, the major manifold in Fig.~\ref{Fig:CeRu4Sn6_DOS1} is Ru-$d$ with traces of Sn-$p$.

\begin{figure}[tp]
 {\includegraphics[width=8cm,clip=,angle=0]{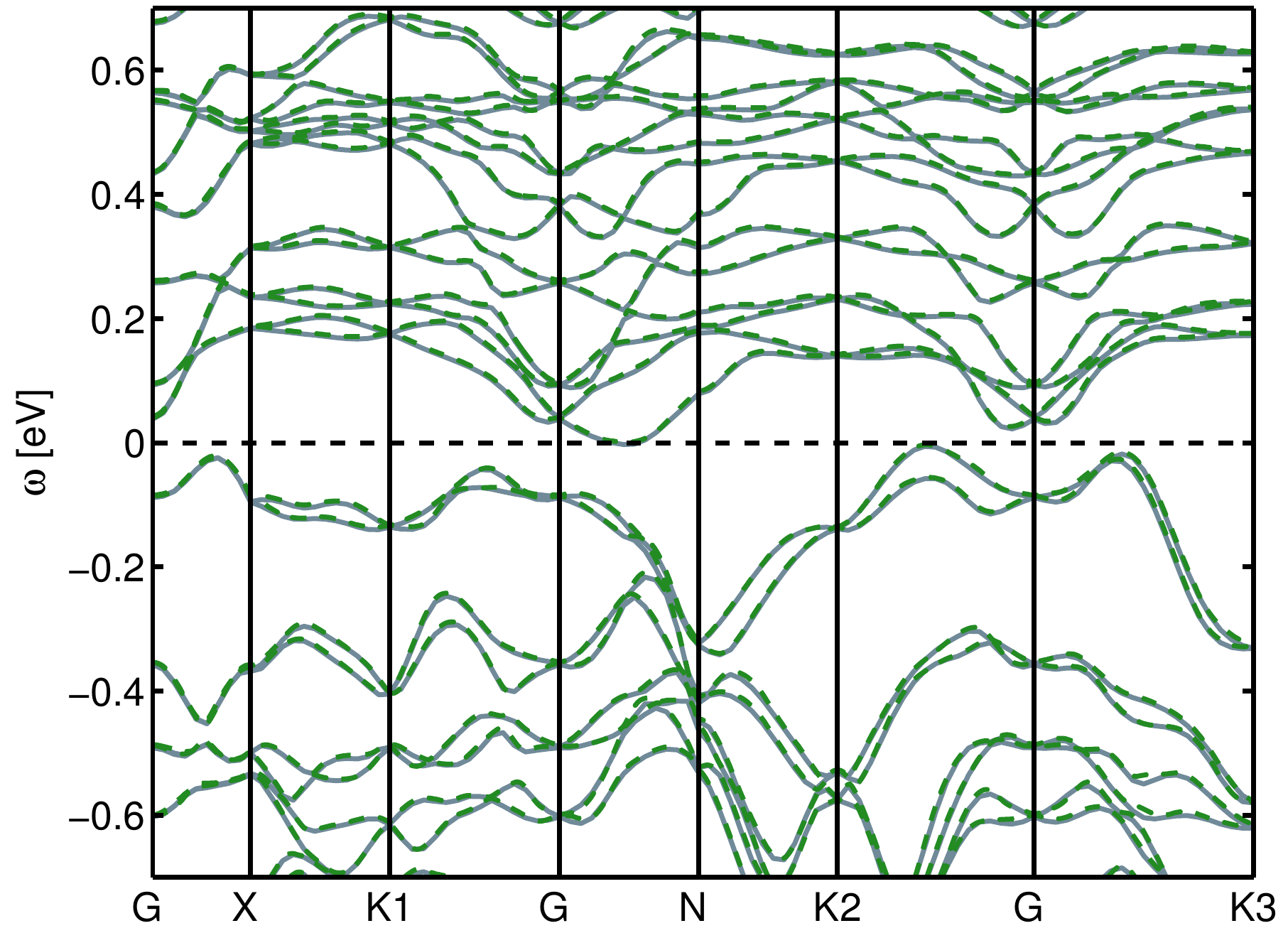}}
     \caption{DFT bandstructure around the Fermi level (solid,gray) perfectly reproduced by the Wannier orbital bands (dashed,green).
   Here and in the following figures, $G =[0\; 0\; 0]$, $X =[0\; 0\;\frac{1}{2}]$, 
   $K1 =[\frac{1}{2}\; 0\; \frac{1}{2}]$,
   $N =[0 \;\frac{1}{2}\; 0]$,
   $K2 =[\frac{1}{2}\; \frac{1}{2}\; 0]$,
   $K3 =[\frac{1}{2}\; \frac{1}{2}\; \frac{1}{2}]$.}
     \label{Fig:CeRu4Sn6_bands1}
 \end{figure}

The corresponding bandstructure of CeRu$_4$Sn$_6$ is shown in Fig~\ref{Fig:CeRu4Sn6_bands1}.  There is
a  small  dip  above the Fermi level. This is due to a DFT direct gap between the 
highest lying occupied band~(mainly Ru-d) and the lowest~(unoccupied) $f$ states with mixed $m_J=\pm 3/2$ and $m_J=\pm 1/2$ character. As one can see in Fig~\ref{Fig:CeRu4Sn6_bands1},  the valence and conduction band both touch the Fermi level however so that the indirect DFT band gap is zero. Also note that  the lack of  inversion symmetry around the Ce atoms lifts the degeneracy between the  $\pm m_J$ components.

\begin{figure}[tp]

\centerline    {$(J=5/2,m_J=+1/2)$}

 {\includegraphics[width=5.cm,clip=,angle=0]{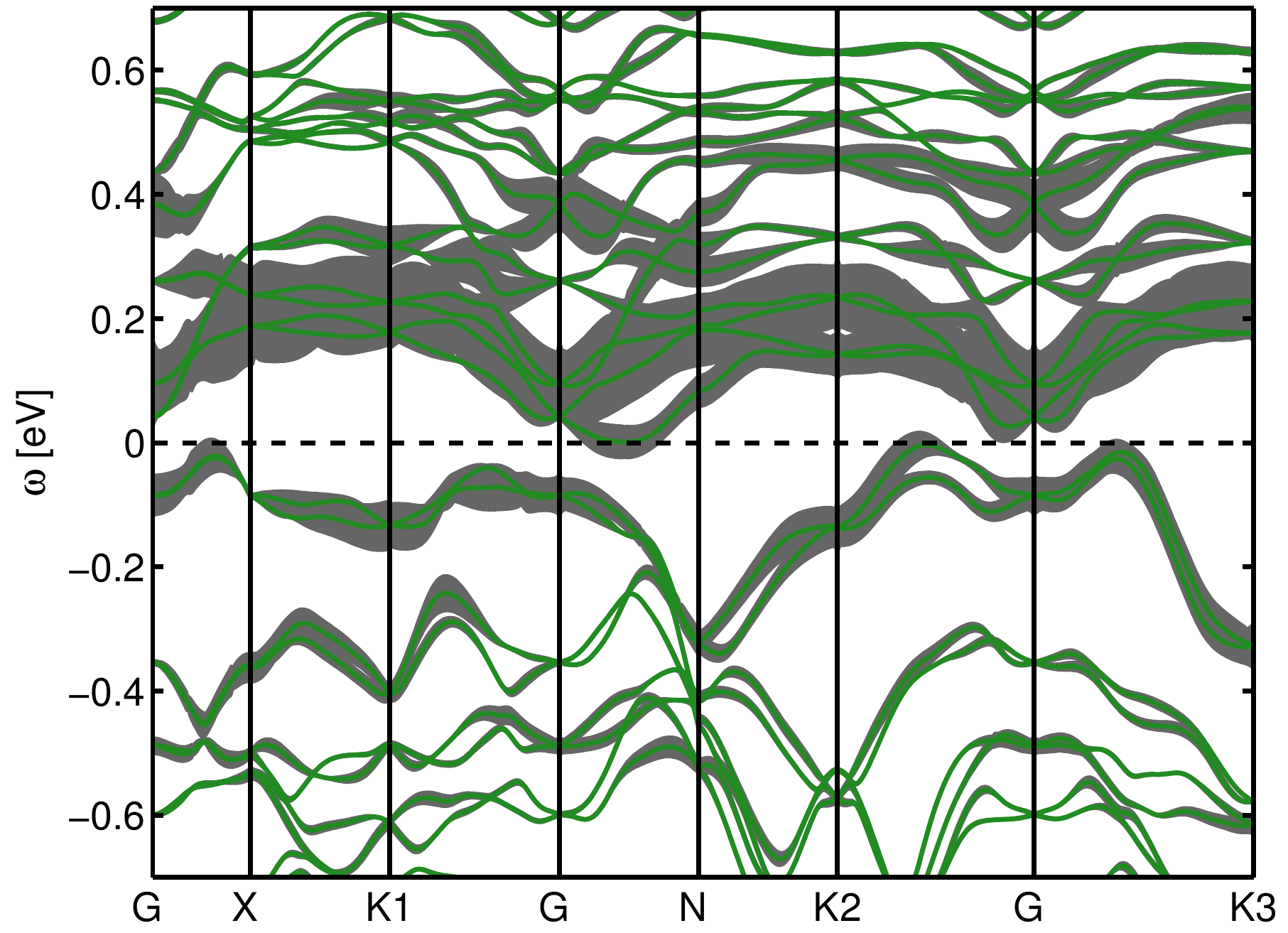}}
 {\includegraphics[width=3.5cm,clip=,angle=0]{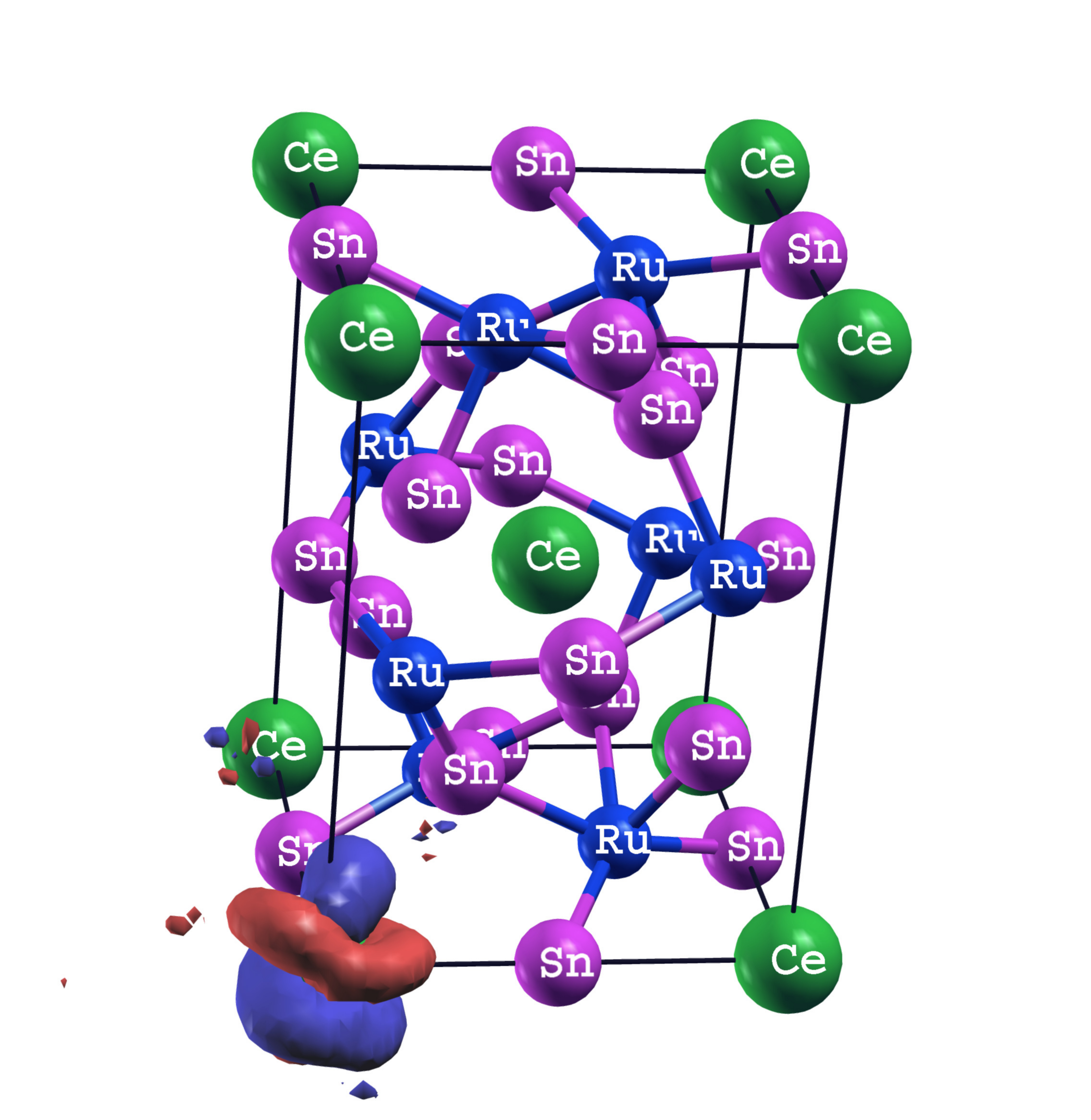}}

\vspace{.5em}

\centerline    {$(J=5/2,m_J=+3/2)$}

{\includegraphics[width=5.cm,clip=,angle=0]{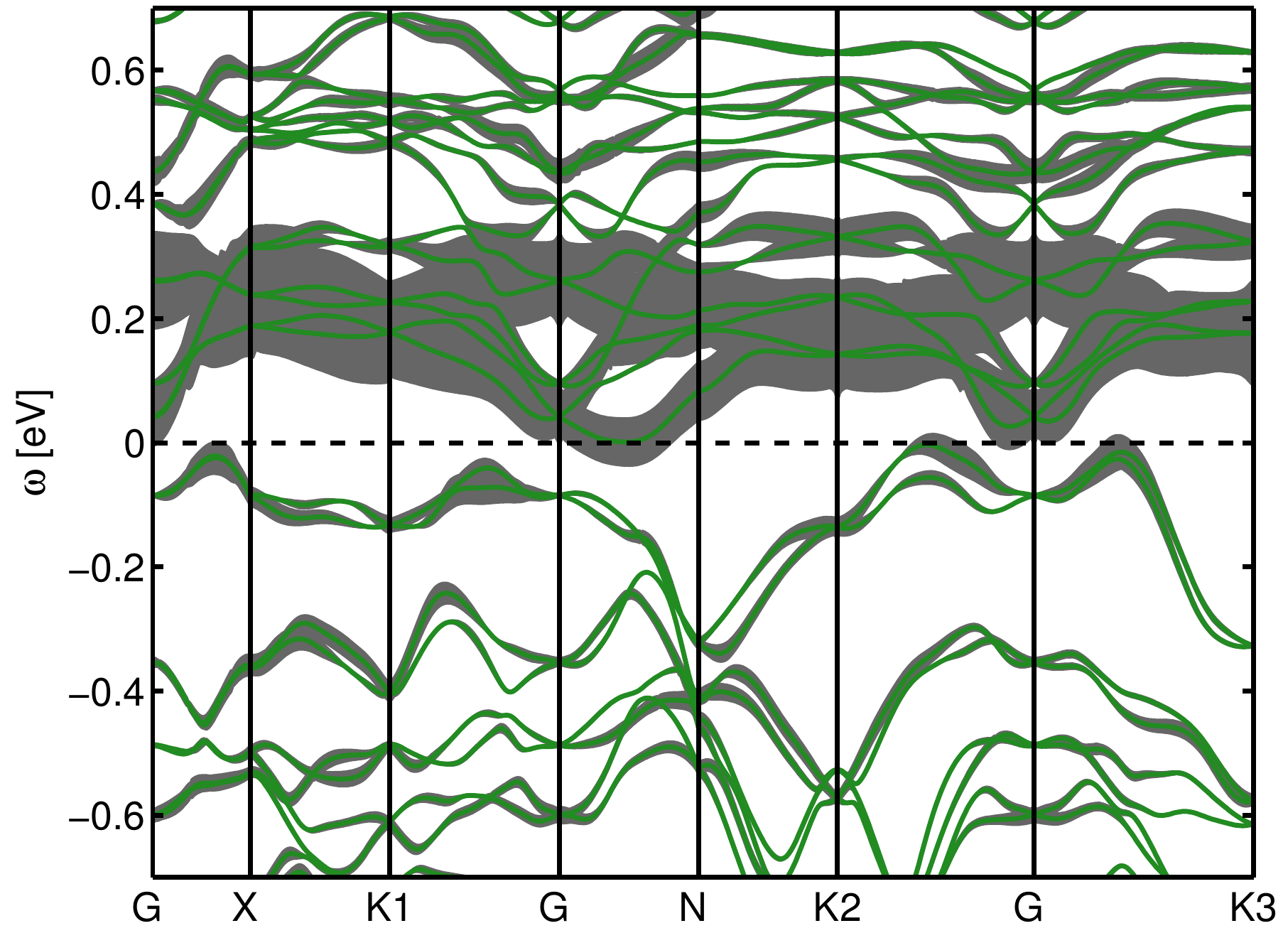}} 
  {\includegraphics[width=3.5cm,clip=,angle=0]{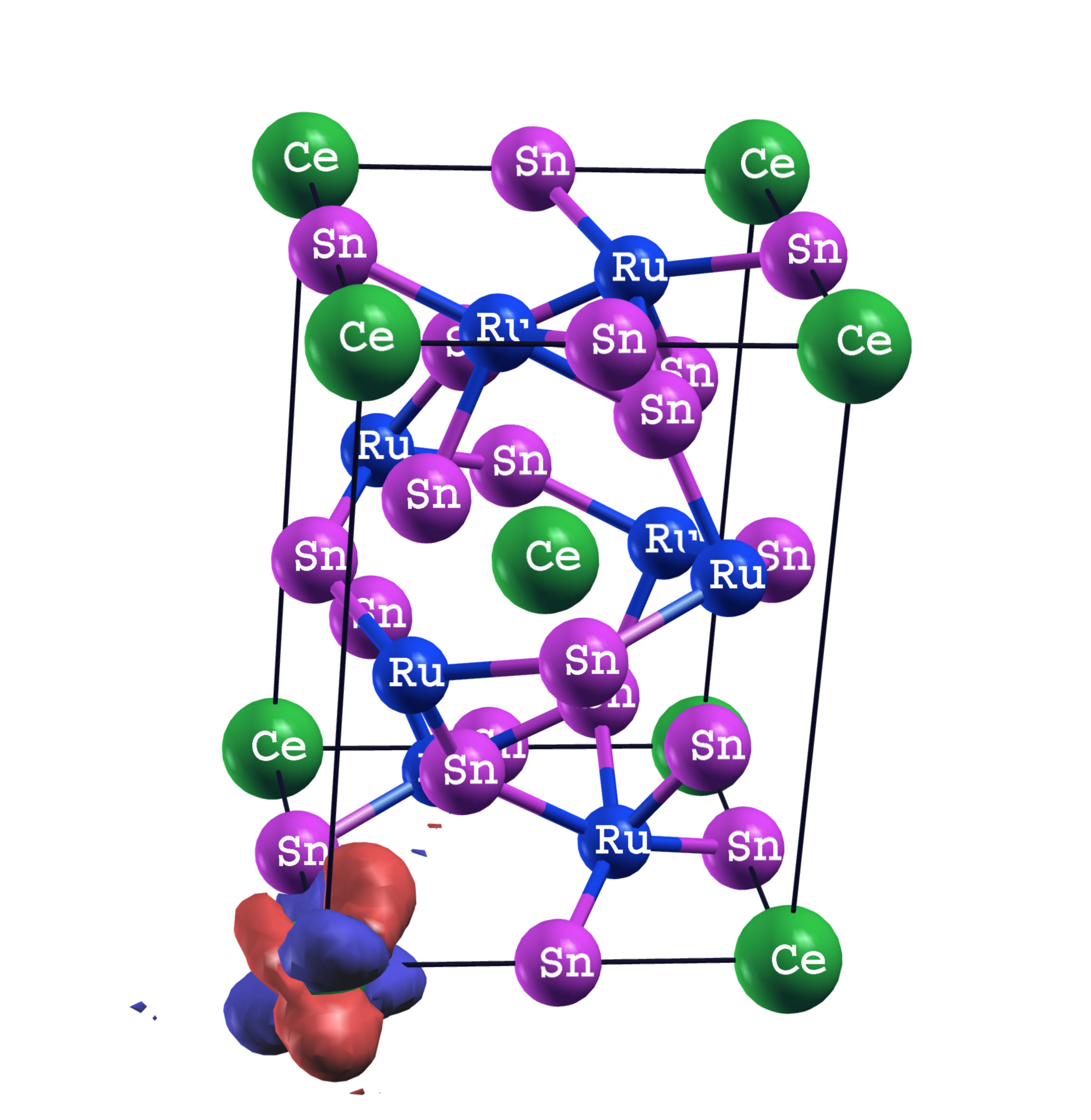}}

\vspace{.5em}

\centerline    {$(J=5/2,m_J=+5/2)$}

{\includegraphics[width=5.cm,clip=,angle=0]{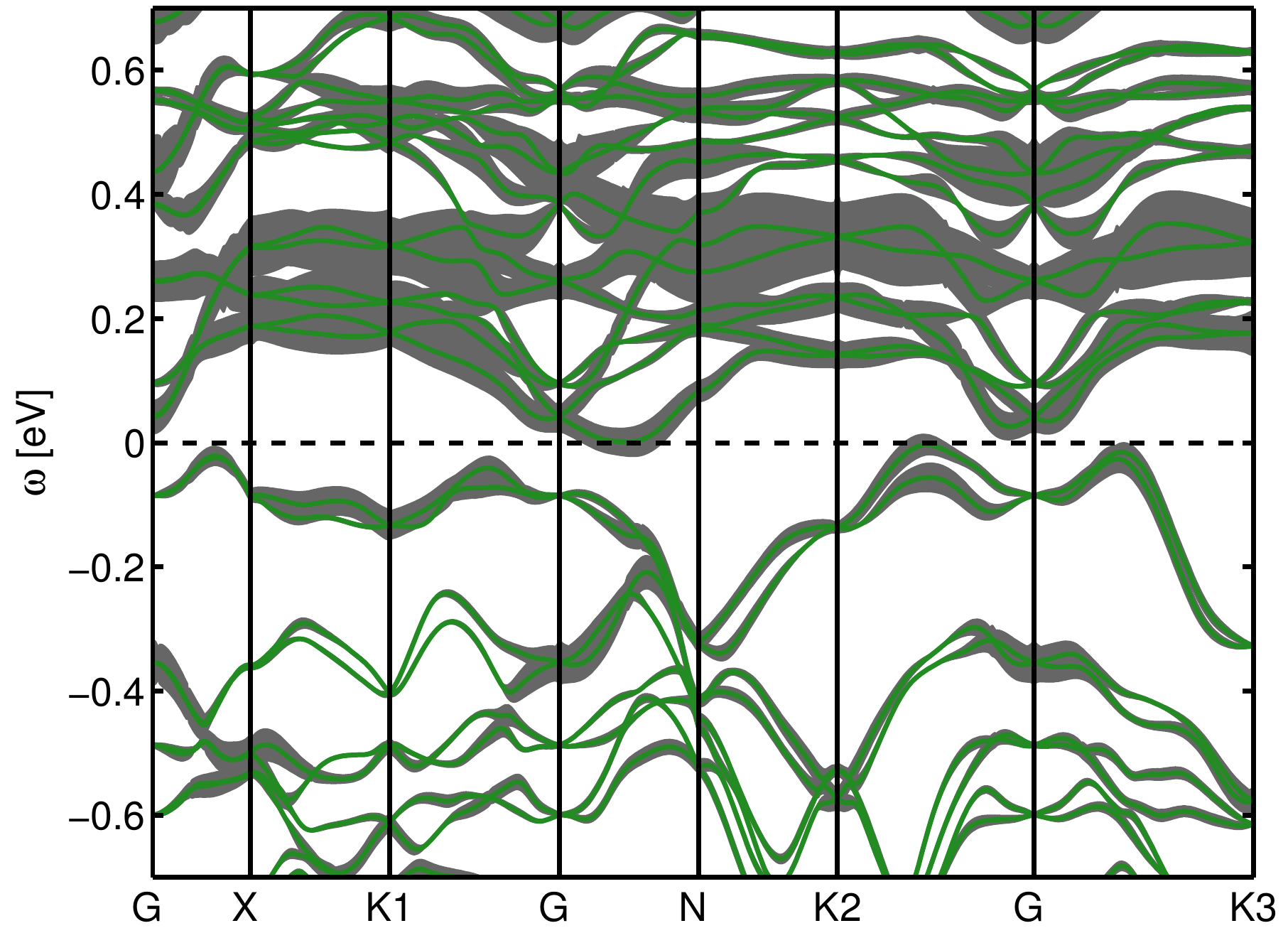}}  
{\includegraphics[width=3.5cm,clip=,angle=0]{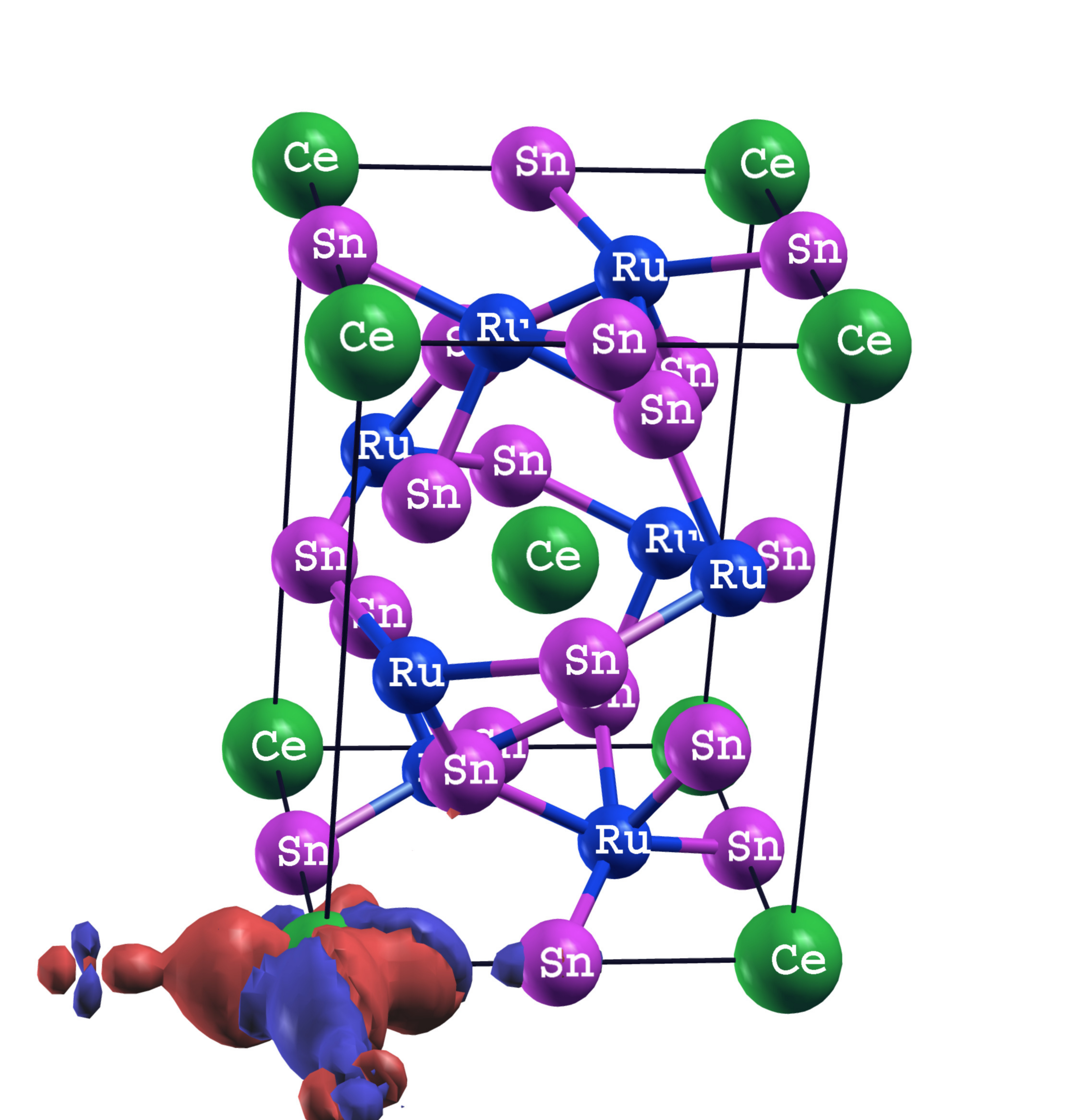}}

    \caption{Wannier orbitals for the Ce-f manifold in CeRu$_4$Sn$_6$ and corresponding fat-band plots. For the charge density plots of the Wannier orbitals, the surface shows $|w(r)|^2=5$ Bohr$^{-3}$ and the color show the sign of the phase $\phi$, i.e. red~(blue) for $\cos(\phi(w(r)))$ positive~(negative).}
    \label{Fig:CeRu4Sn6_P1}
\end{figure}

Next, we do a projection onto maximally localized Wannier orbitals \cite{Marzari1997}  using  the Wien2Wannier \cite{Kunes2010} package.
In this material, the spin-orbit coupling is crucial which makes the identification of adequate Wannier orbitals for modelling the low energy degrees of freedom  more difficult. The decisive aspect is the choice of the atomic orbitals $\phi_n^\uparrow$ and $\phi_n^\downarrow$ for the  Wien2Wannier localization to start with. For example, for  $(J=5/2,m_J=+{3}/{2})$ we start with
\begin{eqnarray}\label{Eq:threehalfinitialstate}
\ket{\phi_{\frac{5}{2},+\frac{3}{2}}}   &\!=\! &-\sqrt{\frac{2}{7}}\ket{\phi_{3,+1}^\uparrow}+\sqrt{\frac{5}{7}}\ket{\phi_{3,+2}^\downarrow}
\end{eqnarray}
where  $\ket{\phi_{3,+1}^\uparrow}$ and $\ket{\phi_{3,+2}^\downarrow}$ denote the
$\ket{L=3,m_L=+1,S=1/2,m_s=+1/2}$ and $\ket{L=3,m_L=+2,S=-1/2,m_s=+1/2}$ in the (lm)-basis, respectively; the prefactors are the  Clebsch-Gordan coefficients.
Besides the Ce-$f$ orbitals, we included all $40$ Ru-$4d$ and $36$ Sn-$5p$ orbitals in the Wannierization.
 Fig.~\ref{Fig:CeRu4Sn6_P1} shows  the most important Ce-$J=5/2$ orbitals and their (fat-band) contribution 
 to  the bandstructur.
The spread of the Ce-$f$ Wannier orbitals is:
   $\Omega_{\frac{5}{2},\pm\frac{3}{2}} = 0.97 \text{ \AA}^2$, $\Omega_{\frac{5}{2},\pm\frac{1}{2}} = 1.10 \text{ \AA}^2$, $\Omega_{\frac{5}{2},\pm\frac{5}{2}} = 1.83 \text{ \AA}^2$,
$\Omega_{\frac{7}{2},\pm\frac{7}{2}} = 1.93 \text{ \AA}^2$, $\Omega_{\frac{7}{2},\pm\frac{5}{2}} = 0.86 \text{ \AA}^2$, $\Omega_{\frac{7}{2},\pm\frac{3}{2}} = 1.32\text{ \AA}^2$, and $\Omega_{\frac{7}{2},\pm\frac{1}{2}} = 0.99 \text{ \AA}^2$. That of the Ru-$d$ Wannier orbitals ranges from $2.32$ to $5.79$ \AA$^2$, and that of the Sn-$p$ is even larger with  $\Omega$ from 5.89 to $18.14$  \AA$^2$.

\section{DFT+DMFT}
\label{Sec:DMFT}
We supplement the $90\times 90$ Wannier Hamiltonian by a local interaction $U=U'=5.5\,$eV   between all Ce-$J=5/2$ states and subtract the double counting in the fully localized limit \cite{Anisimov1991}. The corresponding Hamiltonian is then solved by
 DMFT with Hirsch-Fye quantum Monte Carlo \cite{Hirsch86a} simulations. Note that the Ce-$J=7/2$ orbitals remain unoccupied well above the Fermi level so that we do not need to include them as interacting in DMFT. For the sake of simplicity we neglect Hund's exchange which is justified since Ce only has about one $f$-electron and employ the same DMFT self energy for $\pm m_J$.

\begin{figure}[tp]
   {\includegraphics[width=7.cm,clip=,angle=0]{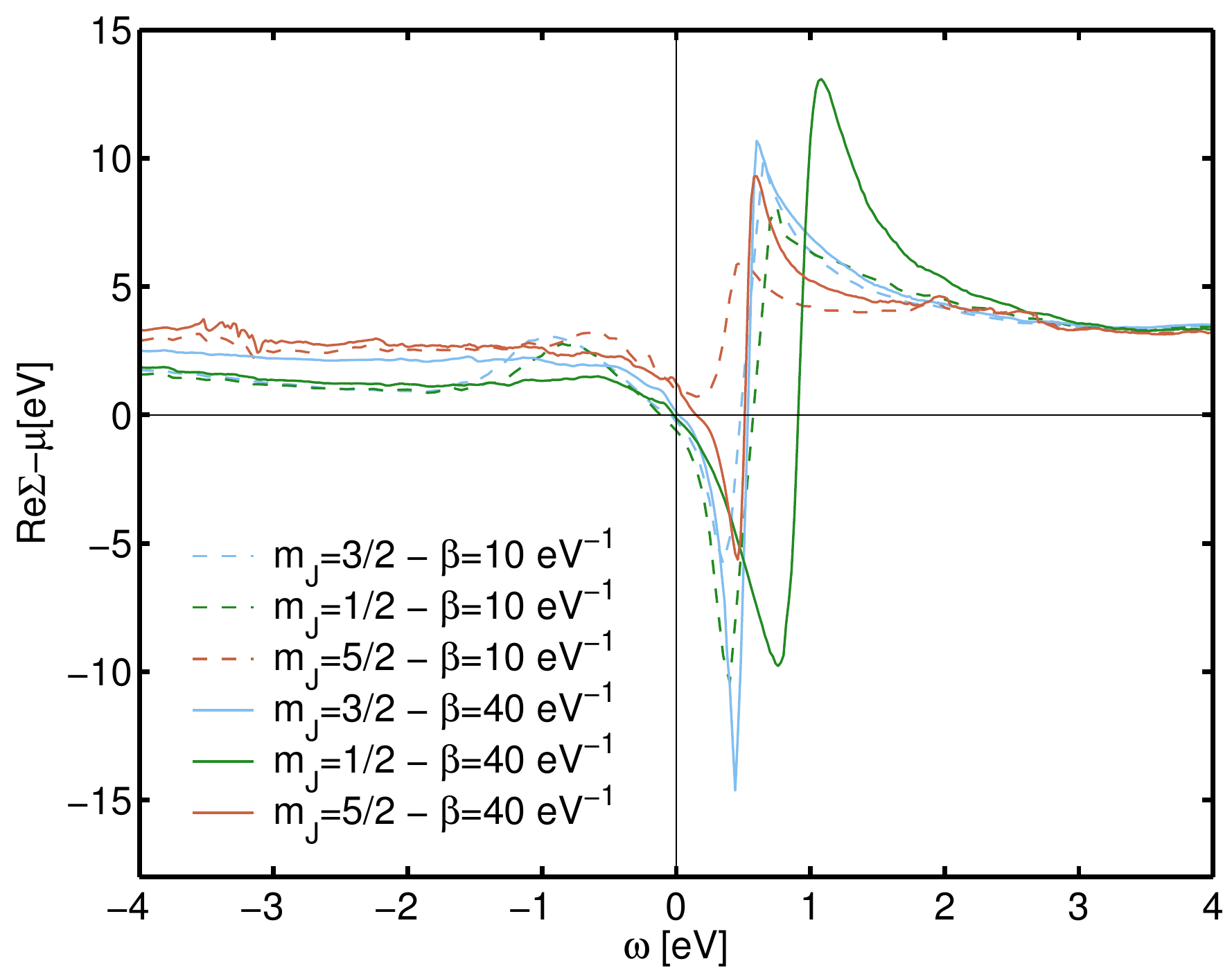}}

   {\includegraphics[width=7.cm,clip=,angle=0]{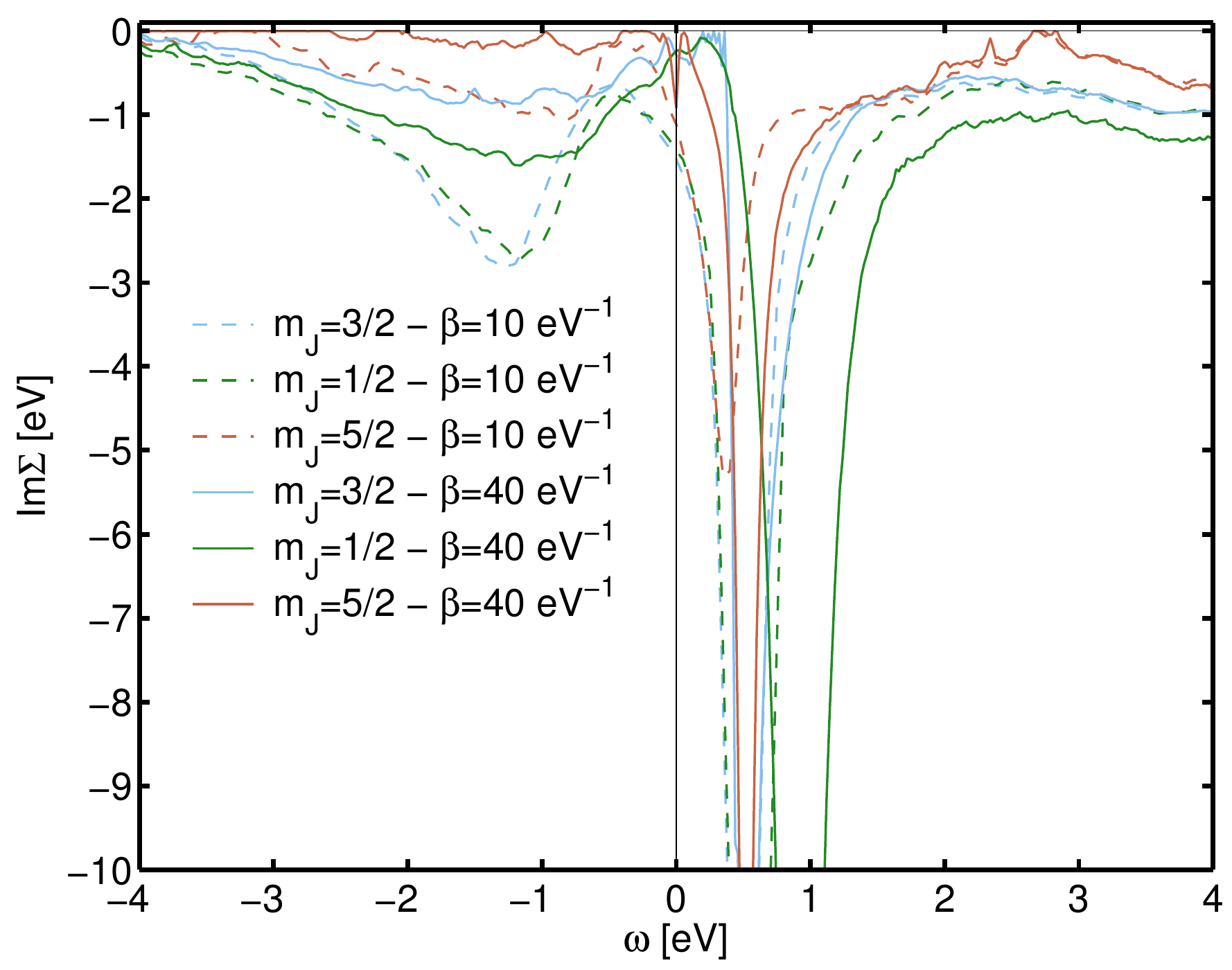}}

      \caption{Orbitally resolved DMFT self energies for real frequencies $\omega$ at inverse temperatures $\beta=10$ eV$^{-1}$ and $40$ eV$^{-1}$. Top: real part; bottom: imaginary part.} 
     \label{Fig:CeRu4Sn6_self1}
 \end{figure}

Let us start our discussion by looking at the orbitally-resolved local electronic densities.
As already mentioned, in DFT the  $|m_J|=1/2$ orbital is slighly more occupied. Specifically the electron density of the 
Ce-${J=5/2}$ manifold is
\begin{eqnarray}\label{Eq:DFTDensities} 
   n_{DFT} &=& (n_{|m_J|=1/2},n_{|m_J|=3/2},n_{|m_J|=5/2})\nonumber \\ 
&=&(0.34\qquad 0.20\qquad 0.28)\\
\end{eqnarray}
Electronic correlations lead with decreasing temperature to  higher  $|m_J|=1/2$ occupations whereas the   $|m_J|=5/2$ states
get depopulated in DMFT:
\begin{eqnarray}\label{Eq:DFTDensities2}
   n_{DMFT}(T=1160\,K) &=& (0.44\quad 0.38\quad 0.15),\\
   n_{DMFT}(T=290\,K) &=& (0.60\quad 0.27\quad 0.08).
\end{eqnarray}
The  higher $|m_J|=1/2$ occupation can be traced back to a reversal of the order of the crystal field levels. As discussed above, in DFT  $|m_J|=3/2$ is the lowest crystal field state, 0.05 eV below $|m_J|=1/2$. To this difference, we need now to add  however the real part of the self energy, which altogether yields the so-called effective crystal field.  The DMFT values (subtracting $\mu$) for the three $m_J$ states are, also see  Fig.~\ref{Fig:CeRu4Sn6_self1}:
\begin{eqnarray}\label{Eq:DFTDensities3}
   \Re\Sigma(T=1160\,K)-\mu &=&  (-0.61\; -0.26\; 0.91 ) \text{ eV},\\
   \Re\Sigma(T=290\,K)-\mu &=& (-0.13\; \phantom{-}0.13\; 1.09)\text{ eV}.
\end{eqnarray}
That is the  $|m_J|=1/2$ is shifted down and eventually becomes
 the lowest effective crystal field state in DFT+DMFT. We attribute this to the larger hybridization between $|m_J|=1/2$ and conduction electrons as well as to the Kondo effect (spin-flipping $m_J=\pm 1/2$ to  $m_J=\mp 1/2$ is possible for conduction electrons with  angular momentum 1/2).

\begin{figure}[tb!]
  \includegraphics[width=8cm]{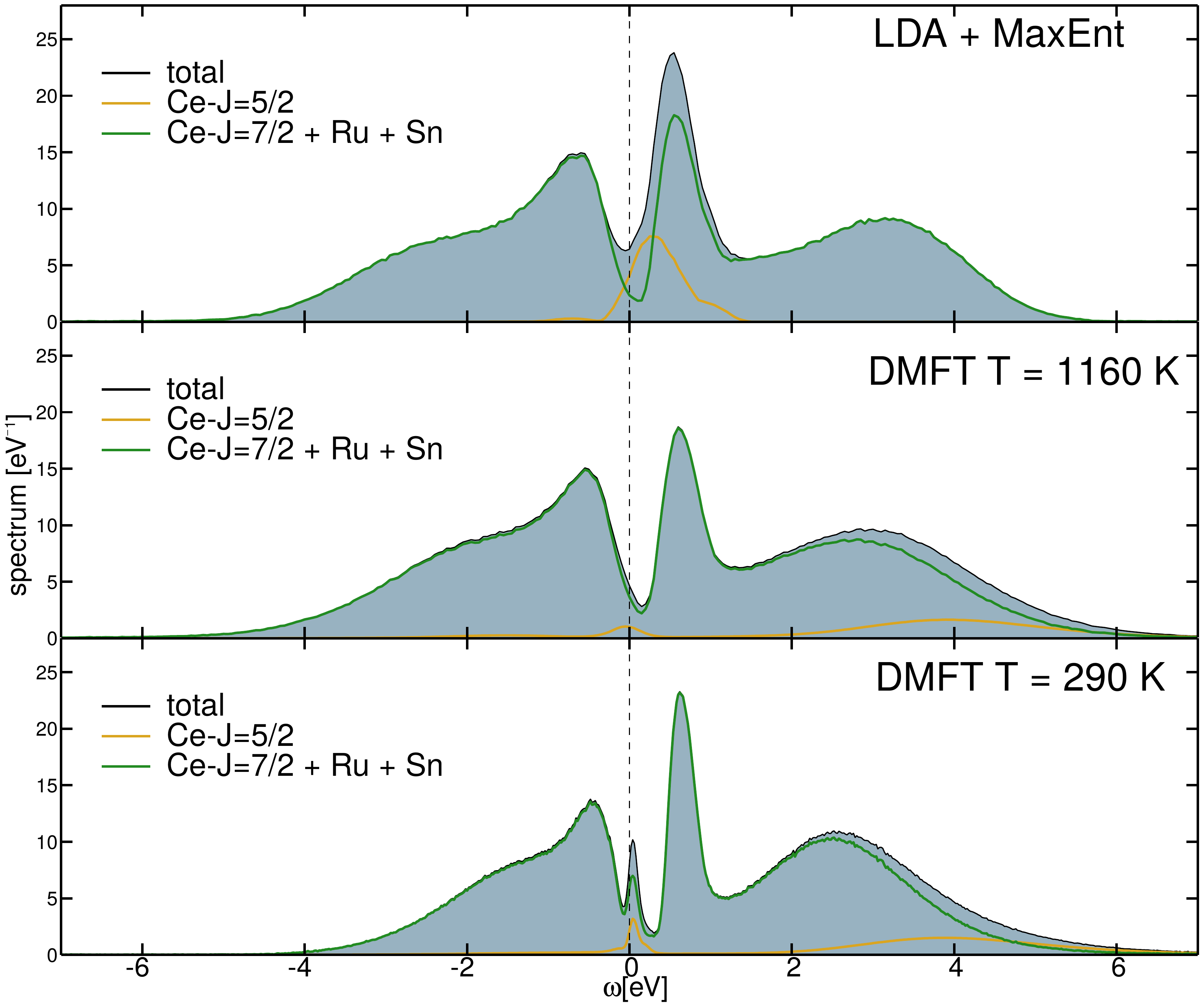}
    \caption{DMFT spectra obtained by the maximum entropy method for $T=1160$ K (middle) and $290$ K (bottom) compared to DFT (top).}
    \label{Fig:CeRu4Sn6_spectrum1}
\end{figure}

\begin{figure}[tb]

   {\includegraphics[width=6.5cm]{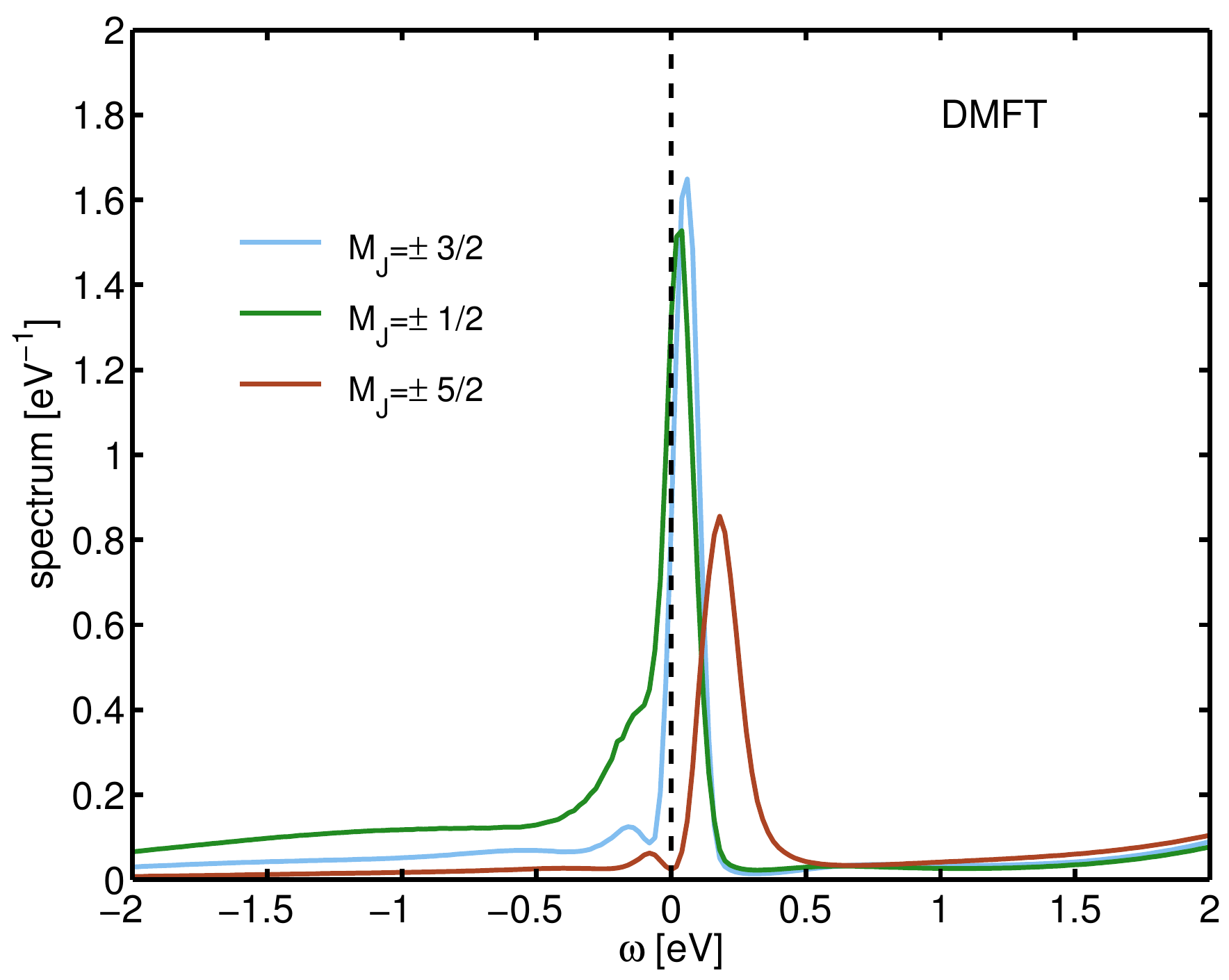}}

   {\includegraphics[width=6.5cm]{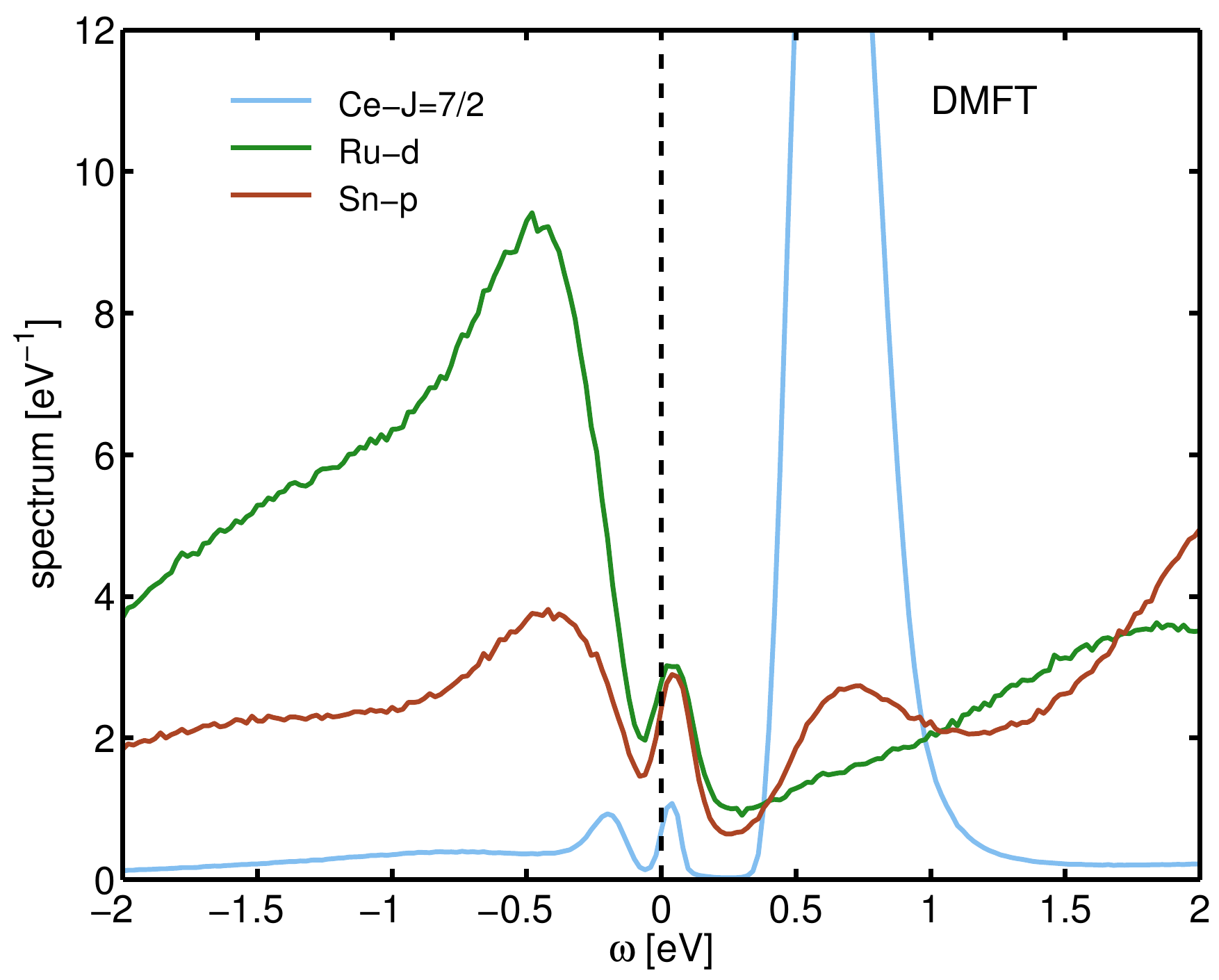}}

     \caption{DMFT spectra at $T=290$ K resolved for the  $J=5/2$ orbitals (left) and corresponding contributions from the other manifolds~(right).}
     \label{Fig:CeRu4Sn6_spectrum2}
 \end{figure}

The Kondo effect can be identified by the development of a Kondo resonance at the Fermi level when decreasing the temperature from  $T=1160$ K  to $290$ K, see  Fig.~\ref{Fig:CeRu4Sn6_spectrum1}.  Fig.~\ref{Fig:CeRu4Sn6_spectrum2} shows a zoom in and an orbital breakup. Clearly, the Kondo resonance at the Fermi level is made up from   $m_J=\mp 1/2$ and   $m_J=\mp 3/2$ states; the  $m_J=\mp 5/2$ states are at a somewhat higher energy due to the (correlation enhanced) crystal field splitting. 

\begin{figure}[tb]
 {\includegraphics[width=8.5cm]{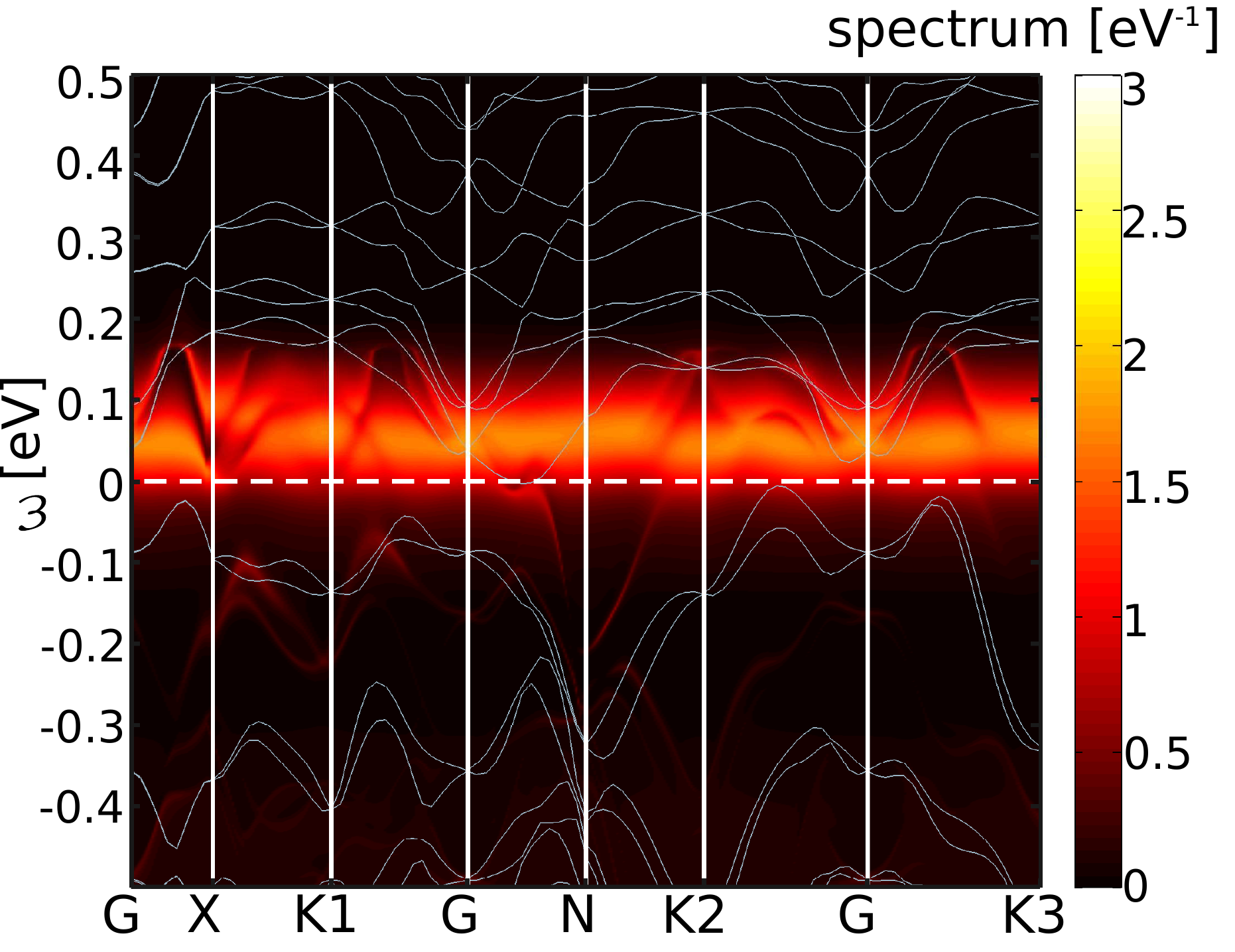}}

{\includegraphics[width=8.5cm]{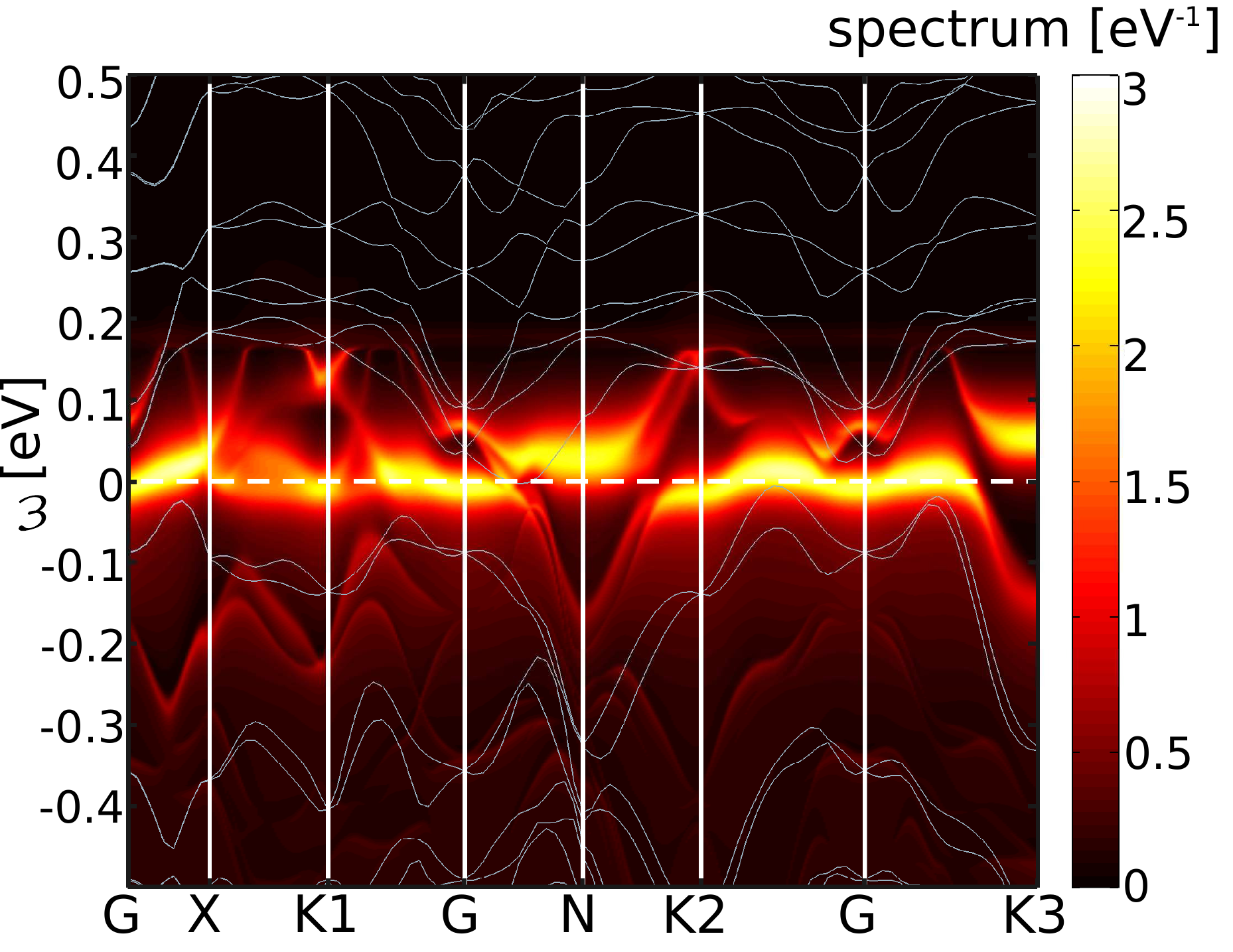}}

    \caption{DMFT $k$-resolved spectra of the two most important orbitals, the $J=5/2,m_J=3/2$ (top) and $J=5/2,m_J=1/2$ (bottom) orbital, at T=290 K. White lines: DFT bandstructure for comparison.
    \label{Fig:CeRu4Sn6_spectrum4}}
\end{figure}

For CeRu$_4$Sn$_6$, there has been a long debate over the nature of the gap and its anisotropy. Fig.~\ref{Fig:CeRu4Sn6_spectrum4} shows the DFT+DMFT spectrum of the two most relevant orbitals, $|m_J|=1/2$ and $|m_J|=3/2$ of the Ce-$J=5/2$ manifold. There is a substantial broadening due to the imaginary part of the self energy which is still huge at 
290K, indicating that we are still considerably above the Kondo temperature.
 
\begin{figure}[tb]
{\includegraphics[width=8.5cm]{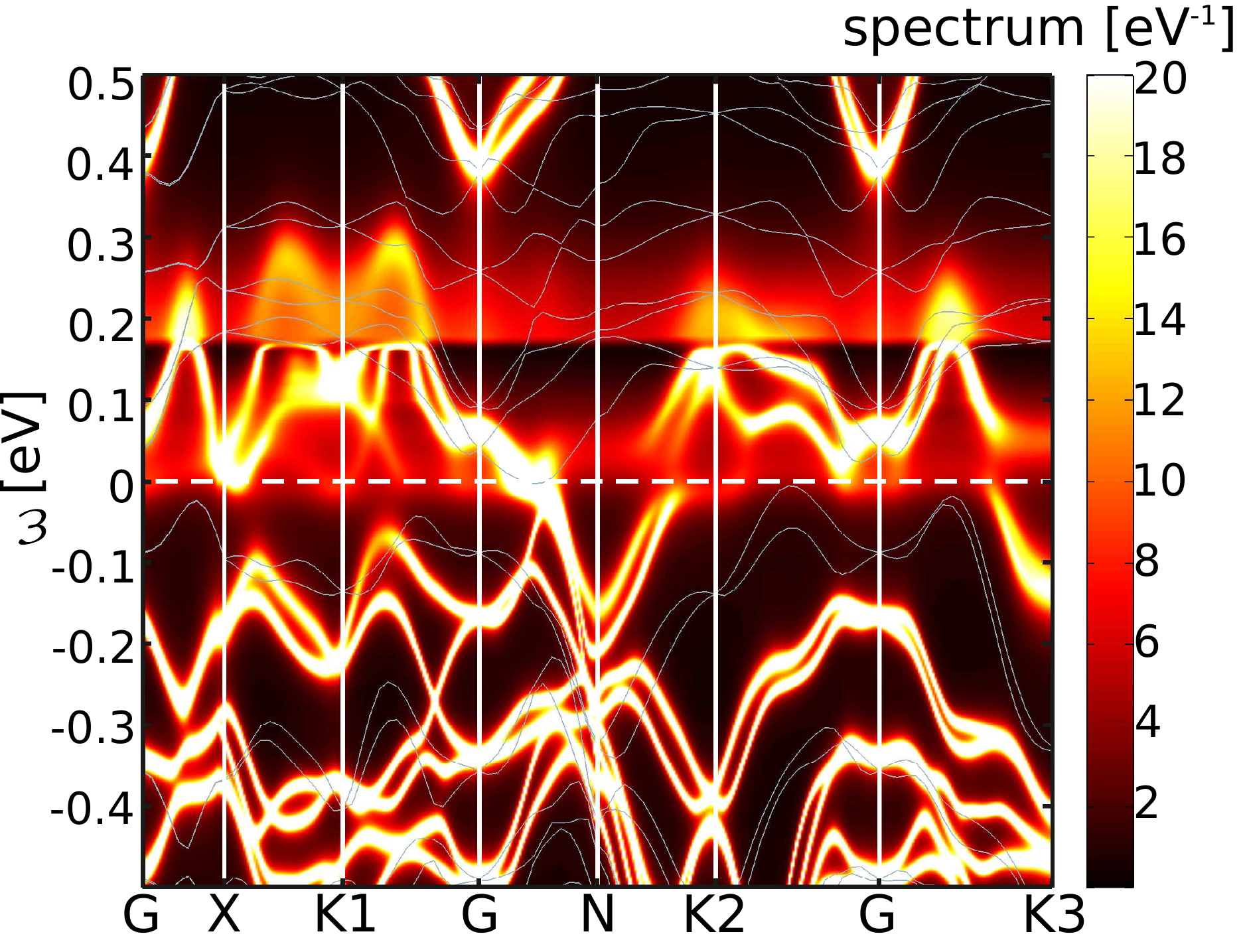}}
{\includegraphics[width=8.5cm]{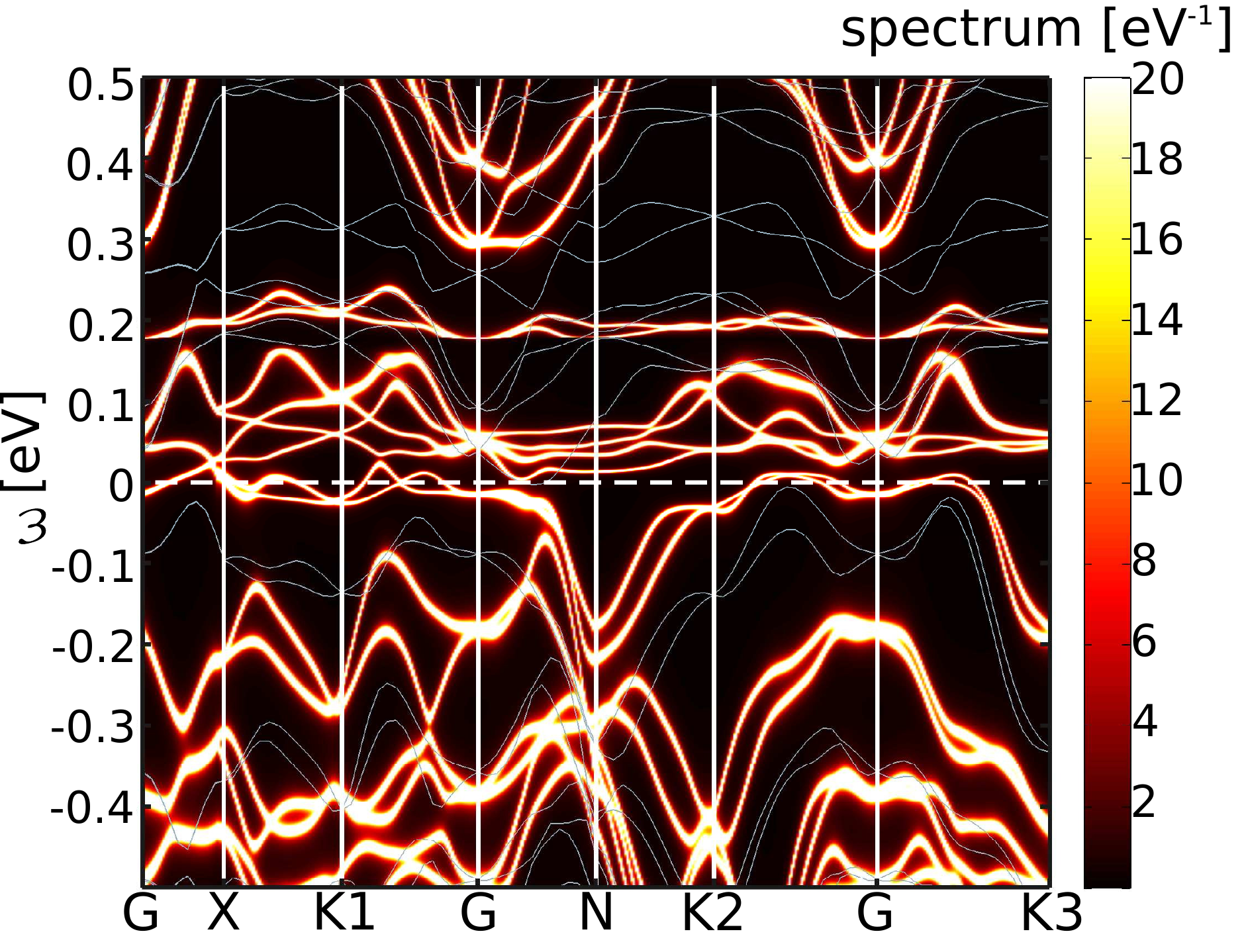}}

    \caption{DMFT $k$-resolved spectrum (top) and that for an artificial $\Sigma(\omega)=\Re(\Sigma_{T=290 K}{\rm DMFT}(\omega))+i0.005$[eV] (bottom). White lines: DFT bandstructure for comparison.}
    \label{Fig:CeRu4Sn6_spectrum3}
\end{figure}

To get a better impression of the correlated bandstructure, we hence set the imaginary part of the self energy to (essentially zero) in  Fig.~\ref{Fig:CeRu4Sn6_spectrum3}. This way one can also mimick the low temperature behavior since the imaginary part of the self energy vanishes like $\sim T^2$  around the Fermi level. One should keep in mind however, that the further build-up of the Kondo resoance might still reshuffle the bandstructure at lower temperatures. 

 Given the small crystal field splitting  between  $|m_J|=1/2$ and $|m_J|=3/2$ and its reversal by correlations as well as the much higher Kondo temperature for a SU(4) than a SU(2) Kondo effect, we can expect that both states will still contribute to the low temperature Kondo resonance.

In principle, Fig.~\ref{Fig:CeRu4Sn6_spectrum3} shows a Kondo insulator with a gap through most of the Brillouin zone. However,  (i) along the $z$ direction from $G$ to $X$  there  a crossing of bands and hence no gap. 
Also in the other directions,  there is (ii) a negative  {\em indirect} band gap. These are two independent meachisms of why  CeRu$_4$Sn$_6$ remains metallic with quite some anisotropy.

The origin of the gap is the usual  Kondo insulator scenario, i.e., the hybridization between Ce-$f$ and   Ru-$4d$(Sn-$5p$) conduction bands. This can be seen by looking at the $N$ and $K_8$ point  in
Fig.~\ref{Fig:CeRu4Sn6_spectrum3} where there is a minimum of a (spin-split)  conduction band around -0.2$\,$eV of mainly Ru-$4d$ character. Around the Fermi level these band go over smoothly into a flat predominantly  Ce $|m_J|=1/2$ band. In principle the same picture can also be found above the Fermi level. However, here it is somewhat less clear, since the conduction band maxima e.g. between $K_8$ and $G$ or $K_3$ and $G$ are not well separated from the other bands. 

Around $X$ and $K_3$, the whole low-energy bandstructure is shifted upwards. Here the gap is well above the Fermi level.
Hence, we only have a direct band gap at most $k$-points whereas the  indirect gap is zero, we have a Kondo semimetal. Note that the term semimetal has been used before, see e.g. Refs.\ \cite{Sato96,Kalvius97,Yartys02}, to classify CeNiSn but to the best of our knowledge in another connotation, i.e., to denote the weak metallic behavior --  even with the Ikeda-Miyake scenario \cite{Ike96.1} in mind, i.e., a hybridization gap with nodes and hence a crossing of conduction and $f$ band.

Moreover, along the $z$ direction, we even have a gap-less crossing of bands, see Fig.~\ref{Fig:CeRu4Sn6_spectrum3}. From  Fig.\ \ref{Fig:CeRu4Sn6_spectrum4} we infer that at 
$G$ the lowest (spin-split) band is $|m_J|=1/2$, the second one $|m_J|=3/2$ and the topmost the predominately Ru conduction band. Hence, this crossing is 
between  $|m_J|=1/2$ and $|m_J|=3/2$. 

Altogether this scenario is much more complicated than a simple model with flat $f$-band. In Ref.\  \onlinecite{Ike96.1} it was concluded that since  $|m_J|=1/2$ orbitals have no  nodes, the  $|m_J|=1/2$ Kondo insulator has to be gapped throughout the Brillouin zone. This scenario is however too simple for
CeRu$_4$Sn$_6$. First of all, there are both, $|m_J|=1/2$ and $|m_J|=3/2$,
bands in the vicinity of the Fermi level; and we  (i) hence observe a crossing of bands. Even if we had no such crossing, we would (ii) nontheless  still
have a Kondo semimetal with remnant metallicity instead of a Kondo insulator since the indirect band gap is negative.  Because of these two reasons, the  idealized modelling of Ref.\
 \onlinecite{Ike96.1} falls short of describing  CeRu$_4$Sn$_6$.

\section{Conclusion}
\label{Sec:conclusion}

In perspective of new x-ray absorption experiments \cite{Severing}
we have presented further  DFT+DMFT results  for   CeRu$_4$Sn$_6$.
In agreement with experiment, the dominat Ce orbital is  $J=5/2$, $|m_J|=1/2$ 
which becomes the lowest crystal field level (and gets more occupied) when the Kondo effect sets in. However, there is also a secondary  $J=5/2$, $|m_J|=3/2$ contribution to the Kondo resonance at the Fermi level. Because we have these two levels there  is  a band crossing along the $z$ direction. Besides, 
 the indirect Kondo gap is negative. That is even without band crossing
we would still have a Kondo semimetal. Both mechanism on their own
yield a remnant metallicity and anisotropy.

\section*{Acknowledgments}

We acknowledge helpful discussions with S. Paschen, P. Thunstr{\"o}m, V. Guritanu, H. Winkler and A. Severing. This work was supported in part by the Austrian Science Fund through the SFB ViCoM F4103-N13 and by
the European Research Council under the European Union’s Seventh Framework Program FP7/ERC through grant agreement n. 306447. The numerical calculations were performed on the Vienna Scientific Cluster (VSC).

\end{document}